\documentclass[english,aip, numerical,jcp,reprint,twocolumn,floatfix]{revtex4-1}

\newcommand{\Ref}[1]{Ref.~\onlinecite{#1}}
\newcommand{\Refs}[1]{Refs.~\onlinecite{#1}}
\usepackage{graphicx}% Include figure files
\usepackage{dcolumn}% Align table columns on decimal point
\usepackage{bm}% bold math
\usepackage{amsmath}
\usepackage{amsfonts}
\begin{document}

\preprint{AIP/123-QED}

\title{
Machine Learning Modeling of  Wigner Intracule Functionals\\
for Two Electrons in One-Dimension
}
\date{\today}     

\author{Rutvij Bhavsar}
\affiliation{Department of Physics, Indian Institute of Technology Kanpur, Kanpur 208016, India}
\affiliation{Tata Institute of Fundamental Research, Centre for Interdisciplinary Sciences, Hyderabad 500107, India}

\author{Raghunathan Ramakrishnan}
\email{ramakrishnan@tifrh.res.in}
\affiliation{Tata Institute of Fundamental Research, Centre for Interdisciplinary Sciences, Hyderabad 500107, India}

%\keywords{two-density matrix, Wigner distribution function}

\begin{abstract}
In principle, many-electron correlation energy can be precisely computed from a reduced Wigner distribution function ($\mathcal{W}$) thanks to a universal functional transformation ($\mathcal{F}$), whose formal existence is akin to that of the exchange-correlation functional in density functional theory. While the exact dependence of $\mathcal{F}$ on $\mathcal{W}$ is unknown, a few approximate parametric models have been proposed in the past. Here, for a dataset of 923 one-dimensional external potentials with two interacting electrons, we apply machine learning to model $\mathcal{F}$ within the kernel {\it Ansatz}. We deal with over-fitting of the kernel to a specific region of phase-space by a one-step regularization not depending on any hyperparameters. Reference correlation energies have been computed by performing exact and Hartree--Fock calculations using discrete variable representation. The resulting models require  $\mathcal{W}$ calculated at the Hartree--Fock level as input while yielding monotonous decay in the predicted correlation energies of new molecules reaching sub-chemical accuracy with  training. 
\end{abstract}

\maketitle
%===INTRO===
\section{Introduction}
The pursuit of reaching chemical accuracy (which is 1 kcal/mol $=0.0434$ eV) in  first-principles predictions of atomic and molecular energetics is persistent. Although the first landmark paper~\cite{hylleraas1928ea} in molecular quantum mechanics~\cite{schaefer2012quantum,hylleraas1964schrodinger} by Hylleraas, had shown how to accurately calculate the energy of the most straightforward non-trivial electronic system, helium atom; to date, achieving this feat for an arbitrary system is far from being reached. The key challenge lies in the incorporation of many-body correlation in the wavefunction---widely used quantum chemistry hierarchies exhibit very strong speed-to-accuracy trade-off, prohibiting accurate modeling of such moderate-sized systems as benzene~\cite{szabo1996modern}. 
To exemplify, only in the last few years, it has become possible to predict the vibrational spectrum of formaldehyde\cite{RaghuRauhut2015}, by incorporating anharmonic effects and agree with experimental measurements to within $\approx$1 cm$^{-1}$. On the other hand, Kohn--Sham density functional theory (KS-DFT)~\cite{parr1989density}, within its domain of applicability, is so successful because its computational complexity is less than or similar to that of Hartree--Fock (HF) while its accuracy often exceeding that of even post-HF methods~\cite{jensen2017introduction}. It is for this very reason, KS-DFT has found broad applicability in various domains such as catalysis, materials design, and even in ab-initio molecular dynamics simulations of protein-ligand complexes. However, reaching chemical
accuracies for energetics using KS-DFT has been a longstanding problem.

In the past decades, several research groups have explored
a variety of {\it non-standard} quantum mechanical methods~\cite{popelier2011solving},
ranging from intracule functional theory (IFT), density matrix renormalization group (DMRG),
reduced-density matrix functional approach (aka 2-RDM method), Sturmian method, quantum Monte-Carlo (QMC), etc. Among these, the IFT is the only method to have been included in a popular quantum chemistry package based on the Gaussian basis set framework with performance benchmarked for the energetics of small molecules~\cite{popelier2011solving,gill2006family}. One of the most attractive features of IFT is that the central variable here is the so-called Wigner intracule, which is a reduced-Wigner distribution function expressed in 
pairwise relative distances and momenta. In the following, we denote the Wigner intracule using the symbol $\mathcal{W}$. 

In analogy with the formal existence of an exact exchange-correlation (XC) functional in DFT that maps the one-electron reduced-density uniquely to a system's ground state energy\cite{hohenberg1964inhomogeneous}, IFT seeks a functional, ${\mathcal F}$,  that maps the ${\mathcal W}$ to the correlation energy, $E_c$~\cite{gill2006family}. A very remarkable feature of this formalism is that the input variable, $\mathcal{W}$, which is formally related to the pair-density, may be approximately coming from a HF wavefunction. Gill {\it et al.}~\cite{o2005benchmark} have shown a simple Gaussian form of ${\mathcal F}$ depending on 2-4 parameters to predict correlation energies of 18 atoms and 56 small molecules rather well. In \Ref{popelier2011solving}, Gill had proposed strategies also to account for static correlation, so as to enhance the method's performance for unsaturated systems. In short, this avenue, to quote Gill {\it et al.}, \cite{gill2011intracule} ``{\it is a fertile, but largely unexplored, middle ground between the simplicity of DFT and the complexity of many-electron wavefunction theories}''.

Be it DFT or IFT, the ultimate goal of finding a universal functional forecasting energetics within the aforementioned chemical accuracy would continue to remain elusive for the foreseeable future. For both problems, the main hurdle is that we do not know how to design the exact universal functional systematically, and to date, XC functionals have been developed mostly via empirical tuning. 
One of the more recent attempts to finding an exact XC functional have utilized kernel-ridge-regression, a machine learning (ML) method, and resulted in a model for $N$ noninteracting spin-less fermions in a one-dimensional   box~\cite{snyder2012finding}. Such data-driven 
modeling strategies, that need not be universally applicable but tailored for given dataset/domain, have now-a-days been applied to a multitude of problems such as quantum mechanical properties of 
molecules~\cite{NN4B3LYP_Chen2003,NN4B3LYP_Chen2004,SVM4CBS_Lomakina2011,Alan_OLED2015,ramakrishnan2015many,RuppPRL2012,hansen2010discrete,faber2017prediction,huang2016communication,ramakrishnan2015electronic,hansen2015machine,ramakrishnan2015big} and extended systems~\cite{GrossMLCrystals2014,ML4Crystals_Wolverton2014,ML4Polymers_Rampi2013,GhiringhelliSchefflerDescriptor_PRL2015,faber2016machine,faber2015crystal}. For more comprehensive reviews, see \Refs{ramakrishnan2017machine,von2017quantum}.

The present study aims to apply ML to discover a numerically exact ${\mathcal F}$ for the IFT. In this first proof-of-concept study, we restrict our explorations to a dataset comprising of one-dimensional atoms and molecules, 
with two electrons. The total number of model systems considered is 923 which 
includes all atoms and molecules that can be
formed by taking up to six atoms with atomic numbers $\le 6$.
To use as training data, we calculate numerically exact $E_c$ using discrete variable representation.

The rest of the paper reads as follows. In the next section, we develop the IFT formalism that is suitable for model studies in one-dimension (1D). We also discuss the electronic Schr\"odinger equation, the composition of the dataset, and details of ML. Then, we present the prediction errors of the models along with the shapes of numerical ${\mathcal F}$ predicted by ML in a data-driven fashion. Finally, we conclude.

\begin{figure}[hpbt!] 
\centering          
\includegraphics[width=8.8cm, angle=0.0]{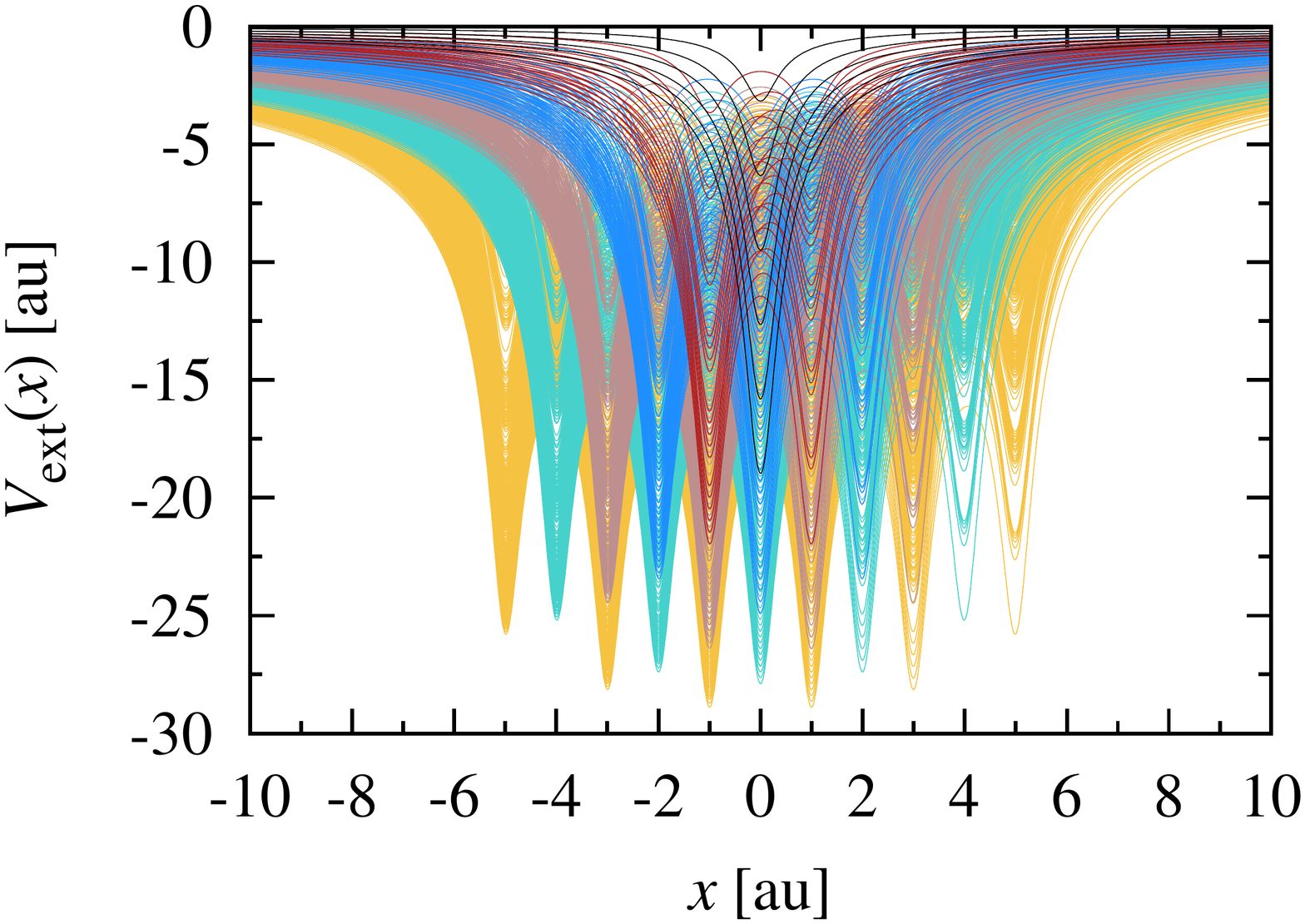}                                       
\caption{Plots of 923 one-dimensional external potentials employed in this study.}
\label{fig:potential}
\end{figure}

%===METHODS===
\section{Methods}
\subsection{Wigner Intracules in One-Dimension}
The postulates of quantum mechanics state the existence of a Hermitian operator $\hat{S}$ for any physically measurable quantity $S$, and that the 
expectation value of $S$ can be calculated for a given wavefunction, in either position ($x$) or momentum ($p$) representation, via the integral
\begin{eqnarray}
\langle \hat{S} \rangle & = & \int dq\, \psi^*(x) \hat{S}(x,p) \psi (x) \nonumber \\
                  & = & \int dp\, \phi^* (p) \hat{S}(x,p) \phi (p)
\end{eqnarray}
The phase-space formulation of quantum mechanics~\cite{kim1991phase} enables calculation of the same expectation value
directly using a scalar observable $S(x,p)$ corresponding to the operator  $\hat{S}$, by employing a distribution function, $f(x,p)$.
\begin{eqnarray}
\langle \hat{S} \rangle & = & \int dx\, f(x,p) s(x,p)
\end{eqnarray}

To date, for practical applications, the most popular choice of $f$ has been the Wigner distribution function (WDF), $W(q,p)$~\cite{wigner1932quantum}. 
In 1D, {\it i.e.} when $q$ is scalar valued, the WDF is defined as
\begin{eqnarray}
W(x,p) & = & \frac{1}{2 \pi} \int_{-\infty}^{+\infty} dy\,  \psi^* \left( x+\frac{\hbar y}{2}\right) \psi \left( x - \frac{\hbar y}{2}\right) e^{i py } \nonumber \\
& &
\end{eqnarray}
This function has also been widely considered to provide a suitable framework to describe the dynamics of interacting quantum systems~\cite{kim1991phase,wlodarz1988note,dahl1982wigner}. 
Despite its widespread popularity, WDF has been known to have counter-intuitive features. First of all, $W(x,p)$ can be negative valued; hence it is a quasi-probability distribution function. Here, without delving further into the epistemological interpretations of WDF, for which we refer the readers to the excellent exposition by Zurek~\cite{zurek2001sub}, we proceed to the derivation of an analytic formula of $\mathcal{W}$ starting from a WDF.

For two electrons confined in 1D, the WDF is defined as the 2D integration
%\begin{widetext}
\begin{eqnarray}
W({\bf x},{\bf p}) & = &
\frac{1}  { 4 \pi^2 }
\int_{-\infty}^{+\infty}  d {\bf y} \, 
\Psi^*\left( {\bf x} + \frac{\hbar {\bf y} }{2} \right)
  \Psi\left( {\bf x} - \frac{\hbar {\bf y} }{2} \right)
e^{
 i  {\bf p} \cdot {\bf y} 
} \nonumber \\
& & 
\end{eqnarray}
%\end{widetext}
Note that $W({\bf x},{\bf p})\in {\mathbb{R}}$, ${\bf x}=\left\lbrace x_1, x_2 \right\rbrace$ and ${\bf p}=\left\lbrace p_1, p_2 \right\rbrace$
being the coordinate and momentum vectors of the two electrons. 
Further, WDF satisfies the normalization convention
\begin{eqnarray}
\int_{-\infty}^{+\infty}  d {\bf x} \,  \int_{-\infty}^{+\infty}  d {\bf p} \, W({\bf x},{\bf p}) & = & 1
\end{eqnarray}
and is symmetric with respect to the exchange of position and momentum variables.
%\begin{widetext}
\begin{eqnarray}
W({\bf x},{\bf p}) & = &
\frac{1} { 4 \pi^2 }
\int_{-\infty}^{+\infty}  d {\bf q} \, 
\Phi^*\left( {\bf p} + \frac{\hbar {\bf q} }{2} \right)
  \Phi\left( {\bf p} - \frac{\hbar {\bf q} }{2} \right)
e^{
 i {\bf x} \cdot {\bf q} 
} \nonumber \\
& &
\end{eqnarray}
%\end{widetext}
From $W({\bf x},{\bf p})$, an intracule that is translationally and rotationally invariant in phase-space can be derived as 
%\begin{widetext}
\begin{eqnarray}
\mathcal{W}(u,v) & = &
\int_{-\infty}^{+\infty}  d {\bf x} \,  \int_{-\infty}^{+\infty}  d {\bf p} \, 
W({\bf x},{\bf p})  \nonumber \\
& &
\delta \left( |x_1-x_2| - u \right)
\delta \left( |p_1-p_2| - v \right)
\label{eq:one}
\end{eqnarray}
%\end{widetext}

In principle, this 4D integration can be numerically performed, albeit using a tiny integration step that is required to capture the effect of Dirac-$\delta$ functions. To arrive at an expression that is amenable to quick computation, the explicit dependence of Eq.\ref{eq:one} on the $\delta$-functions has to be eliminated. To this end, we invoke the  relation, $\delta(|a| - b)=\delta(a - b)+\delta(-a - b)$, resulting in
%\begin{widetext}
\begin{eqnarray}
& & \mathcal{W}(u,v) = \int_{-\infty}^{+\infty}  d {\bf x} \,  \int_{-\infty}^{+\infty}  d {\bf p} \, 
W({\bf x},{\bf p})   \\
& & \left[ \delta \left( x_1-x_2 - u \right) + \delta \left( x_1-x_2 + u \right) \right]  \nonumber \\
& & \left[ \delta \left( p_1-p_2 - v \right) + \delta \left( p_1-p_2 + v \right) \right], \nonumber
\end{eqnarray}
where we have utilized the fact that $\delta(x)$ is an even function.
We can now invoke the elementary properties of $\delta$-function and arrive at an expression with four displaced WDFs. 
%\begin{widetext}
\begin{eqnarray}
&& \mathcal{W}(u,v) =  \frac{1}{\pi \hbar} \int_{-\infty}^{+\infty}  d {x_1} \,  \int_{-\infty}^{+\infty}  d {p_1} \,  \nonumber   \\
& &   \left[ W(x_1,p_1,x_1+u,p_1+v) + W(x_1,p_1,x_1-u,p_1+v) + \right. \nonumber \\
& &   \left. W(x_1,p_1,x_1+u,p_1-v) + W(x_1,p_1,x_1-u,p_1-v)   \right] 
\label{eq:three}
\end{eqnarray}
%\end{widetext}
The first of the four terms in Eq.~\ref{eq:three} can be simplified through a straight-forward analytic integration over $p_1$ variable  
%\begin{widetext}
\begin{eqnarray}
 & & W (x_1,p_1,x_1+u,p_1+v) =
 \frac{1}  { 4 \pi^2 }
 \int_{-\infty}^{+\infty}
d {\bf y} d x_1 d p_1 \, \nonumber \\
& & 
\Psi^*\left( x_1 + \frac{\hbar y_1 }{2},  x_1 + u + \frac{\hbar y_2 }{2}\right)
  \Psi\left( x_1 - \frac{\hbar y_1 }{2},  x_1 + u - \frac{\hbar y_2 }{2}\right) \nonumber \\
& & \exp \left[
 i \left\lbrace  
p_1 y_1 + \left( p_1 + v \right)y_2
\right\rbrace
\right]
\label{eq:twoo}
\end{eqnarray}
%\end{widetext}
Then, by making the substitutions 
$\alpha=\left( y_1 + y_2 \right)/\sqrt[]{2}$ and 
$\beta=\left( y_1 - y_2 \right)/\sqrt[]{2}$ along with
the standard formula
\begin{eqnarray}
 \int_{-\infty}^{+\infty} d p_1 e^{i p_1 \sqrt[]{2} \alpha} = 2 \pi \delta  \left( \sqrt[]{2} \alpha \right),
\end{eqnarray}
we can simplify the RHS of Eq.~\ref{eq:twoo} as a 3D integration.
\begin{widetext}
\begin{eqnarray}
% & = & 
 \frac{1}  {  2 \pi }
 \int_{-\infty}^{+\infty}
d \alpha d \beta d x_1 \, 
\Psi^*\left( x_1 + \frac{ \alpha+\beta }{2\,\sqrt[]{2}},  x_1 + u +\frac{ \alpha - \beta }{2\, \sqrt[]{2}} \right)
  \Psi\left( x_1 - \frac{ \alpha+\beta }{2\, \sqrt[]{2}},  x_1 + u -\frac{ \alpha - \beta }{2 \,\sqrt[]{2}} \right)
e^{
 i v
(\alpha - \beta )/\sqrt[]{2}
}
\delta  \left( \sqrt[]{2} \alpha \right)
\end{eqnarray}
\end{widetext}
This expression, can be further simplified as the following 2D intergration by utilizing the property of $\delta$-function.
\begin{widetext}
\begin{eqnarray}
 \frac{1}  { 2\,\sqrt[]{2} \pi } \int_{-\infty}^{+\infty} d \beta d x_1 \, 
\Psi^*\left( x_1 + \frac{ \beta }{2\, \sqrt[]{2}},  x_1 + u - \frac{ \beta }{2\, \sqrt[]{2}} \right)
  \Psi\left( x_1 - \frac{ \beta }{2\, \sqrt[]{2}},  x_1 + u + \frac{ \beta }{2\, \sqrt[]{2}} \right)
e^{-i v \beta /\sqrt[]{2}}
\end{eqnarray}
\end{widetext}

Finally, by collecting the simplified expressions for all four displaced WDFs in Eq. \ref{eq:three}, we arrive
at the following formula which requires a wavefunction as the argument and 
computes the intracule through a 2D integration
\begin{widetext}
\begin{eqnarray}
\mathcal{W}(u,v)  & = & 
\frac{2}{\pi} \int_{-\infty}^{+\infty}  d \beta d x_1 \, 
 \left[ 
\Psi^*\left( x_1 + \beta, x_1 + u - \beta \right)
\Psi  \left( x_1 - \beta, x_1 + u + \beta \right) +  \right. \nonumber \\
&  & \left. \Psi^*\left( x_1 + \beta, x_1 - u - \beta \right)
\Psi  \left( x_1 - \beta, x_1 - u + \beta \right)
\right]
\cos \left( 2 v \beta\right)
\label{fig:Wuv}
\end{eqnarray}
\end{widetext}
%It is worth noting that while $W({\bf x},{\bf p})$ being the most popular distribution function, 
%and $\mathcal{W}(u,v)$ 

\subsection{Computational Details and Dataset}
In the following, we employ atomic units, {\it i.e.}, mass of electron, $m_e=1$, and $\hbar=1$. 
Within the Born--Oppenheimer approximation, the non-relativistic molecular electronic Hamiltonian for two electrons is given by
\begin{equation}                                                                          
\hat{H}=\sum_{i=1}^{2} \hat{T}(x_i) + \sum_{i=1}^{2} \hat{V}_{\rm ext}(x_i) +         
\sum_{i=1}^{2} \sum_{j>i}^{2} \hat{V}_{ee}(x_i,x_j),                                  
\end{equation}
The one-particle kinetic energy term takes the usual form                                
\begin{equation}  
\hat{T}(x_i) =  -\frac{1}{2} d^2_{x_i}
\end{equation}
with the  external potential  defined as                   
\begin{eqnarray}                                                                          
V_{\rm ext}(x_i) & = & \sum_{A=1}^{M}                                                     
 \frac{-Z_{A}}{\sqrt{(x_i-x_{A})^{2} + \alpha }}.                                      
 \label{eq:potvne}                                                                  
\end{eqnarray}
where $M$ is the number of nuclei,
$Z_A$ is the nuclear charge of atom $A$,  
and $\alpha \ge 0$; 
while the electron-electron                                                            
interaction operator is given by                                                      
\begin{eqnarray}                                                                          
V_{\rm ee}(x_i,x_j) & = &  \frac{1}{\sqrt{(x_i-x_j)^{2} +  \alpha }}.                   
 \label{eq:potvee}                                                                   
\end{eqnarray}
For both attraction and repulsion potential energy operators, with $\alpha=0$ we recover the hard-Coulomb limit, while for other values of $\alpha$, we have soft-Coulomb operators.    
In this study, we have used  $\alpha=1.0$, which results in potential profiles qualitatively similar to that of a hard-Coulomb potential with an incomplete cusp ~\cite{ramakrishnan2012control}. For multi-well systems, we have utilized a separation of 2 bohr. With the resulting Hamiltonians, we have 
performed HF and exact calculations using the basis set free approach, 
discrete variable representation (DVR)~\cite{colbert1992novel}. 
In all calculations, we have used 128 grids in the domain $-15 \le x_1, x_2 \le +15$ bohr. In the 2D DVR calculations this results in matrices of size $128^2=16384$.

It is important to note that soft-Coulomb potentials are not an approximations to exact potentials in 1D. Using Gauss's law it can be shown that the true ({\it i.e.} exact) electrostatic interaction potential can be obtained as the solution of Poisson's equation. In 1D, such a solution amounts to a potential linearly depending on the coordinate (see APPENDIX 1). Hence, the so-called hard-Coulomb potentials with $\alpha=0$ in Eq.~\ref{eq:potvne} and Eq.~\ref{eq:potvee} do not represent exact electrostatic potentials but rather serve as model potentials providing qualitative insights. This latter case is peculiar in its own regards---such interactions lead to diverging energies. For instance, in the case of a 1D analogue of helium atom, corresponding to $Z=2$ in Eq.\ref{eq:potvne}, exact calculations performed with $8192^2=67.1\times10^{6}$ product basis functions yield the ground state energy to be $-2.22420955$, $-5.83766144$, $-14.07977723$, 
$-30.76728906$, and $-60.61204718$ hartree for $\alpha=1,0.1,0.01,0.001$ and $0.0001$, respectively. 
An analytic proof for this divergence is presented separately in APPENDIX 2. 
However, it may be worthwhile to note that the 1D hard-Coulomb problem can be tackled approximately by introducing the Dirichlet boundary condition~\cite{ball2017molecular}.

Our dataset includes all possible 1D atoms and molecules with two electrons. 
The total number of systems is limited
by the maximal number of atoms ($N_{\rm max}$) and the maximal nuclear charge ($Z_{\rm max}$).
In this way, we arrive at $Z_{\rm max}$ single-wells, 
$Z_{\rm max}(Z_{\rm max}+1)/2$ double-wells, 
$Z_{\rm max}(Z_{\rm max}+1)(Z_{\rm max}+2)/3!$ triple-wells,
and in general, 
$Z_{\rm max}(Z_{\rm max}+1)\ldots(Z_{\rm max}+N_{\rm max}-1)/N_{\rm max}!$  potentials
with $N_{\rm max}$ minima. 
Using $N_{\rm max}=Z_{\rm max}=6$ we arrive at 923 1D potentials
that are on display in Fig.~\ref{fig:potential}.

%=======================
\subsection{Intracule-kernel modeling}
%=======================
The exact  Wigner intracule correlation functional ($\mathcal{F}$) provides an 
injective mapping between the 
Wigner intracule derived from a HF wavefunction and the exact many-body correlation
energy: $\mathcal{F}\left[ \mathcal{W}\left({\boldsymbol \theta}\right)\right] = E^{\rm corr}$. Here, 
${\boldsymbol \theta}$ collectively denotes  the intracule 
variables $\left\lbrace u, v\right\rbrace$.
In this work, 
we model $\mathcal{F}$ as a {\it functional transformation} that maps $\mathcal{W}\left({\boldsymbol \theta, V^{\rm ext}}\right)$ to $E^{\rm corr}\left(V^{\rm ext}\right)$, where $V^{\rm ext}$ is the external potential. 
Our goal, in particular, is to model $\mathcal{F}$ as 
a kernel, $\mathcal{G}$. In this case, the mapping is established through the inner product
\begin{eqnarray}
E^{\rm corr}\left(V^{\rm ext}\right)=\int 
{\mathcal G}\left({\boldsymbol \theta}\right)
\mathcal{W}\left({\boldsymbol \theta}\right)d{\boldsymbol \theta}
\end{eqnarray}
While it is not the purpose of this study to investigate the formal 
existence of such a kernel, we assume its existence as in \cite{popelier2011solving,gill2004wigner,gill2006family,gill2011intracule,o2005benchmark}, and aim to
find its numerical approximation, which we denote by $\widetilde{\mathcal{G}}$.

The problem of finding an optimal kernel that minimizes the prediction error in a least-squares 
fashion leads to the unconstrained loss function
\begin{eqnarray}
{\mathcal L} = 
\underset{\widetilde{\mathcal{G}}}{{\rm min}}\sum_{k} \left[ E_{k}^{\rm corr}-\int\widetilde{\mathcal{G}}\left({\boldsymbol \theta}\right)\mathcal{W_{\mathit{k}}}\left({\boldsymbol \theta}\right)d{\boldsymbol \theta} \right]^2
\end{eqnarray}
The same equation, written in matrix-vector notation is given as
\begin{eqnarray}
{\mathcal L} = 
\underset{\widetilde{\mathcal{G}}}{{\rm min}} \left\Vert {\bf E}^{\rm corr}-\int\widetilde{\mathcal{G}}\left({\boldsymbol \theta}\right)\mathcal{{\bf W}}\left({\boldsymbol \theta}\right)d{\boldsymbol \theta}\right\Vert _{2}^2,
\label{eq:19}
\end{eqnarray}
where ${\bf E}^{\rm corr}$ is a column vector of correlation energies of the 
training set, and $\mathcal{{\bf W}}$ is a super-matrix containing intracules of
all the training molecules. The notation $\left\Vert \cdot \right\Vert _{2}$ indicates an $L_2$- or Euclidean-norm. 
While this equation is exactly solvable ensuring zero loss, ${\mathcal L}$, 
like all the rank-deficient system of equations, 
the solution is not unique, and 
one can arrive at one of the infinite solutions all satisfying
\begin{eqnarray}
\widetilde{\mathcal{G}}\left({\boldsymbol \theta}\right) = {\bf W}^{+}{\bf E}^{\rm corr}
\end{eqnarray}
In the above equation,  ${\bf W}^{+}$ is the  Moore--Penrose pseudo-inverse~\cite{moors1920reciprocal,penrose1955generalized} of ${\bf W}$
\begin{eqnarray}
{\bf W}^{+}={\bf W}^{\rm T} \left( {\bf W}{\bf W}^{\rm T} \right)^{-1}
\end{eqnarray}
Furthermore, the main drawback of the intracule functional thus obtained will be the
lack of continuity in ${\boldsymbol \theta}$. In other words, the values
of the kernel over a given range of $u$ and $v$ will tend to oscillate so rapidly that the overall
performance will be governed by an excessive overfitting to the training set.

For such problems, one of the widely used procedures to quench overfitting 
is regularizing the model~\cite{scholkopf2002learning} by constraining the magnitude of the kernel. 
In kernel-ridge-regression (KRR)---that has widely been applied in the ML modeling of properties across chemical space---it is the $L_2$-norm of the coefficient vector that is added as a penalty to the
loss-function~\cite{ramakrishnan2017machine} after multiplying with the Lagrangian multiplier, aka the length-scale hyperparameter. 
In the present study, where our inference is not done via KRR (the kernel in KRR here should not be confused with the intracule kernel functional, $\mathcal{G}$), our goal is  to include a penalty function in Eq.~\ref{eq:19}. For this purpose,
we use the $L_2$-norm of $\widetilde{\mathcal{G}}$. 
\begin{eqnarray}
\mathcal{L} = \underset{\widetilde{\mathcal{G}}}{{\rm min}}
 \left\Vert {\bf E}^{\rm corr}-\int\widetilde{\mathcal{G}}\left({\boldsymbol \theta}\right)\mathcal{{\bf W}}\left({\boldsymbol \theta}\right)d{\boldsymbol \theta}\right\Vert _{2}^2
+ \left\Vert \widetilde{\mathcal{G}}\left({\boldsymbol \theta}\right)\right\Vert_{2}^2
\label{eq:dgelsy}
\end{eqnarray}
It may be of interest to note that, this problem, when employing an $L_1$-norm would be analogous to the 
least-absolute shrinkage and selection operator (LASSO) approach \cite{tibshirani1996regression}, that
has recently been so successfully employed to map the structure of binary materials~\cite{GhiringhelliSchefflerDescriptor_PRL2015}.
While the LASSO approach involves a two-fold non-linear optimization, Eq.~\ref{eq:dgelsy} has the desirable feature of being a convex problem that 
can be solved exactly using pure linear algebra
%via the GELSY procedure~\cite{anderson1999lapack}
resulting in a unique minimum norm solution $\widetilde{\mathcal{G}}$. 
Accordingly, the exact solution to this problem can be obtained as
\begin{eqnarray}
\widetilde{\mathcal{G}}\left({\boldsymbol \theta}\right) = \widetilde{{\bf W}^{+}}{\bf E}^{\rm corr}
\label{eq:22}
\end{eqnarray}
where  $\widetilde{{\bf W}^{+}}$ is the 
pseudoinverse of ${\bf W}$
obtained via rank-revealing QR (RRQR) factorization with column interchanges~\cite{anderson1999lapack,golub2012matrix,gu1996efficient}.
In this procedure, the matrix ${\bf W} \in  \mathbb{R}^{m \times n}$  is first decomposed as
\begin{eqnarray}
{\bf W}
={\bf Q}{\bf R} {\bf P}^{\rm T}
=
{\bf Q}\begin{bmatrix}
    {\bf R}_{11}       & {\bf R}_{12} \\
    0            & {\bf R}_{22} \\
\end{bmatrix}
{\bf P}^{\rm T}
\end{eqnarray}
where ${\bf Q} \in \mathbb{R}^{m \times m}$ is an orthogonal matrix satisfying
${\bf Q}^{\rm T}{\bf Q} = {\bf I}$ and ${\bf R}_{11} \in \mathbb{R} ^{q \times q}$ is an upper diagonal matrix. 
The permutation matrix ${\bf P}$ and the effective rank $q$
are chosen such that ${\bf R}_{11}$ is well-conditioned ({\it i.e.} the condition number, $\kappa$, is smaller than a threshold) and the $L_2$-norm of the matrix ${\bf R}_{22} \in \mathbb{R}^{(m-q) \times (m-q)}$ is numerically negligible
\begin{eqnarray}
{\bf R}  
\approx
\begin{bmatrix}
    {\bf R}_{11}       & {\bf R}_{12} \\
    0            & 0 \\
\end{bmatrix}
\end{eqnarray}
It is possible to further simplify the super-matrix ${\bf R}$ via an orthogonal transformation from the right
to eliminate the off-diagonal matrix ${\bf R}_{12}$ 
\begin{eqnarray}
{\bf R} \approx
{\bf V}^{\rm T}
\begin{bmatrix}
    {\bf R}_{11}       & {\bf R}_{12} \\
    0            & 0 \\
\end{bmatrix}
{\bf V}
=\begin{bmatrix}
    {\bf T}_{11}       & 0 \\
    0            & 0 \\
\end{bmatrix}
{\bf V}
\end{eqnarray}
%===
The overall decomposition can now be expressed as 
\begin{eqnarray}
{\bf W} 
= {\bf Q}
\begin{bmatrix}
    {\bf T}_{11}       & 0 \\
    0            & 0      \\
\end{bmatrix}{\bf V} {\bf P}^{\rm T}
\label{eq:eight}
\end{eqnarray}
where ${\bf T}$ is also a triangular matrix. The 
pseudoinverse of ${\bf W}$ is now given by
\begin{eqnarray}
\widetilde{{\bf W}^{+}} = {\bf P} {\bf V}^{\rm T}\begin{bmatrix}
    {\bf T}_{11}^{-1}    & 0    \\
    0        & 0      \\
\end{bmatrix}{\bf Q}^{\rm T}
\label{eq:nine}
\end{eqnarray}
and is used with Eq.~\ref{eq:22}.

\begin{figure}[hptb!] 
\centering  
\includegraphics[width=8.7cm, angle=0.0]{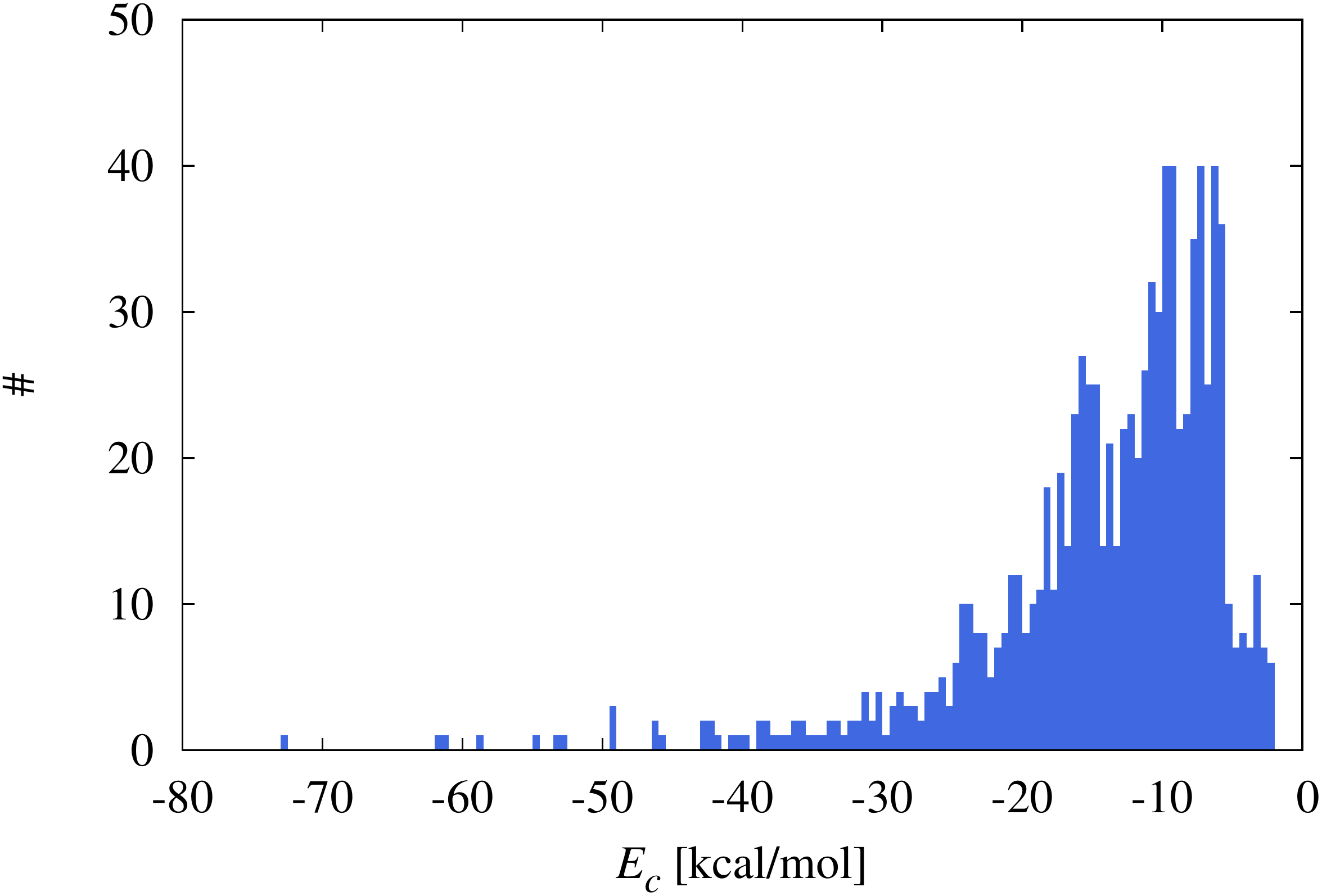}                                       
\caption{
Distribution of correlation energies across 923 one-dimensional systems.
}
\label{fig:ecoor}
\end{figure} 

We note in passing that in the training of the ML models, we have kept the most localized 83 potentials, corresponding to single-, double- until four-wells in the training set because such {\it small} systems are under-represented in the entire dataset. A similar strategy had been employed in an earlier ML study of molecular electronic
properties~\cite{montavon2013machine}. Such pruning, of course, is not necessary when dealing with large training sets as in large scale KRR studies~\cite{ramakrishnan2015big,ramakrishnan2015many,ramakrishnan2015electronic}.
After training, the correlation energy of a new out-of-sample system, that was not used in the
training of the ML model is estimated as
\begin{eqnarray}
E^{\rm corr}\left(V^{\rm ext, new}\right)=\int 
\widetilde{\mathcal G}\left({\boldsymbol \theta}\right)
\mathcal{W}^{\rm new}\left({\boldsymbol \theta}\right)d{\boldsymbol \theta}
\label{eq:pred}
\end{eqnarray}

\section{Results and Discussion}
Our target quantities of interest, which are the correlation energies of the 923 systems, ranges from -2 to -73 kcal/mol. 
The distribution of $E_c$ over the systems is displayed in Fig.~\ref{fig:ecoor}. Majority of the molecules exhibit moderate correlation energies of about 10 kcal/mol. In comparison, the correlation energy of a 3
d Helium atom is -26.4 kcal/mol. The system exhibiting weakest correlation also coincides with that of steepest single-well, the 1D C$^{4+}$ ion, with $E_c=-2.1$ kcal/mol. This trend is understandable because within the soft-Coulomb approximation, the electron-electron interaction is now relatively weaker compared to the steeper external potential, confining the system to a small region.

%===
\begin{figure*}[hpt] 
\centering          
\includegraphics[width=11cm, angle=0.0]{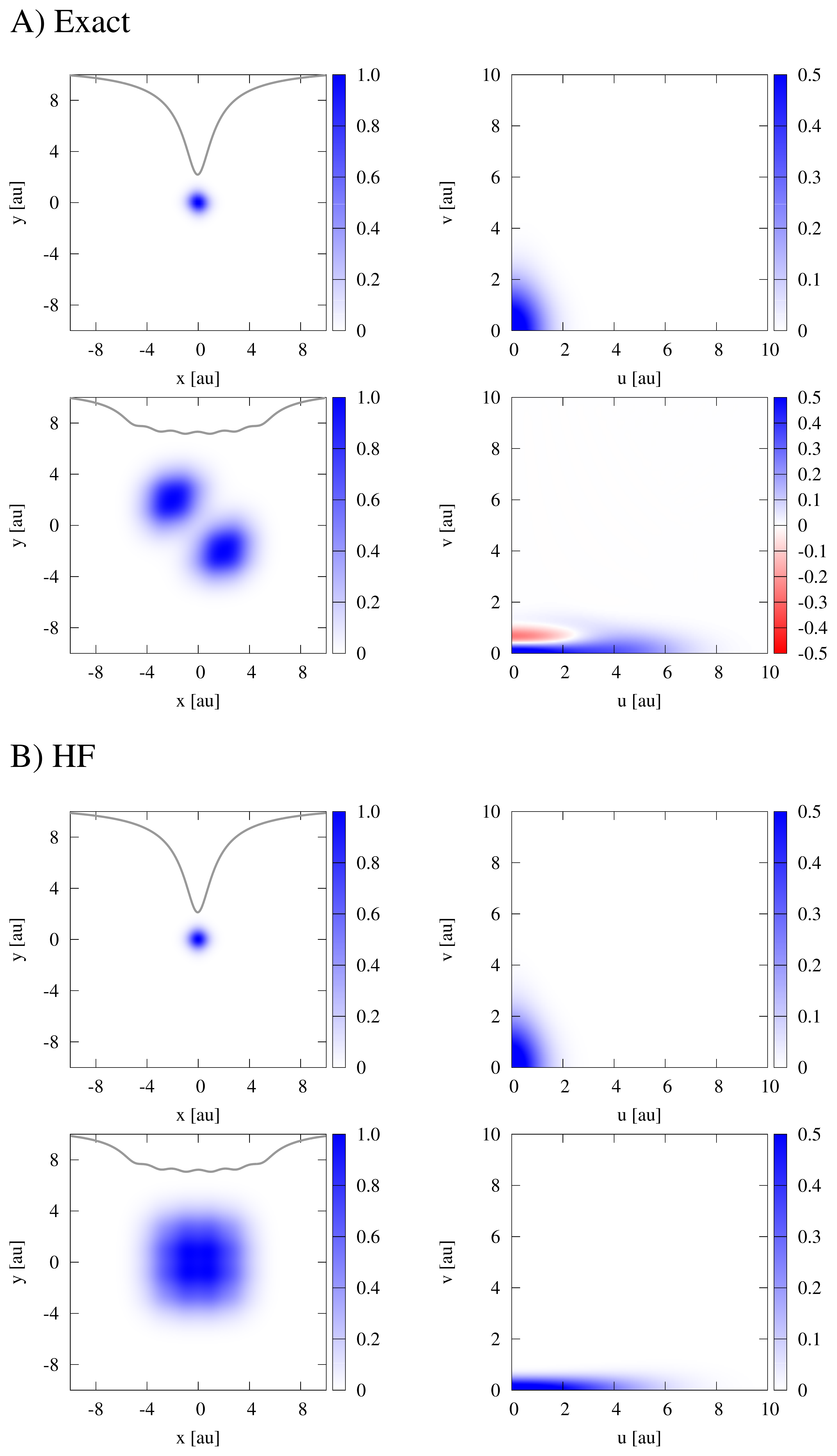}
\caption{ 
Two-electron probability density of the ground state, $\rho(x_1,x_2)$ (LEFT) and the corresponding Wigner-intracules, $W(u,v)$ (RIGHT) are plotted for the most localized (single well, $Z=6$), and most delocalized (six-wells, $Z_i=1$; $i=1,\ldots,6$)
potentials in the dataset: 
A) using the exact ground state wavefunction, and B) using the restricted Hartree--Fock wavefunction. 
The shapes of the external potentials are shown as gray curves. For clarity, probability densities are normalized to maximal values.}
\label{fig:psiW}
\end{figure*}

%===
\begin{figure}[hpt] 
\centering  
\includegraphics[width=8.5cm, angle=0.0]{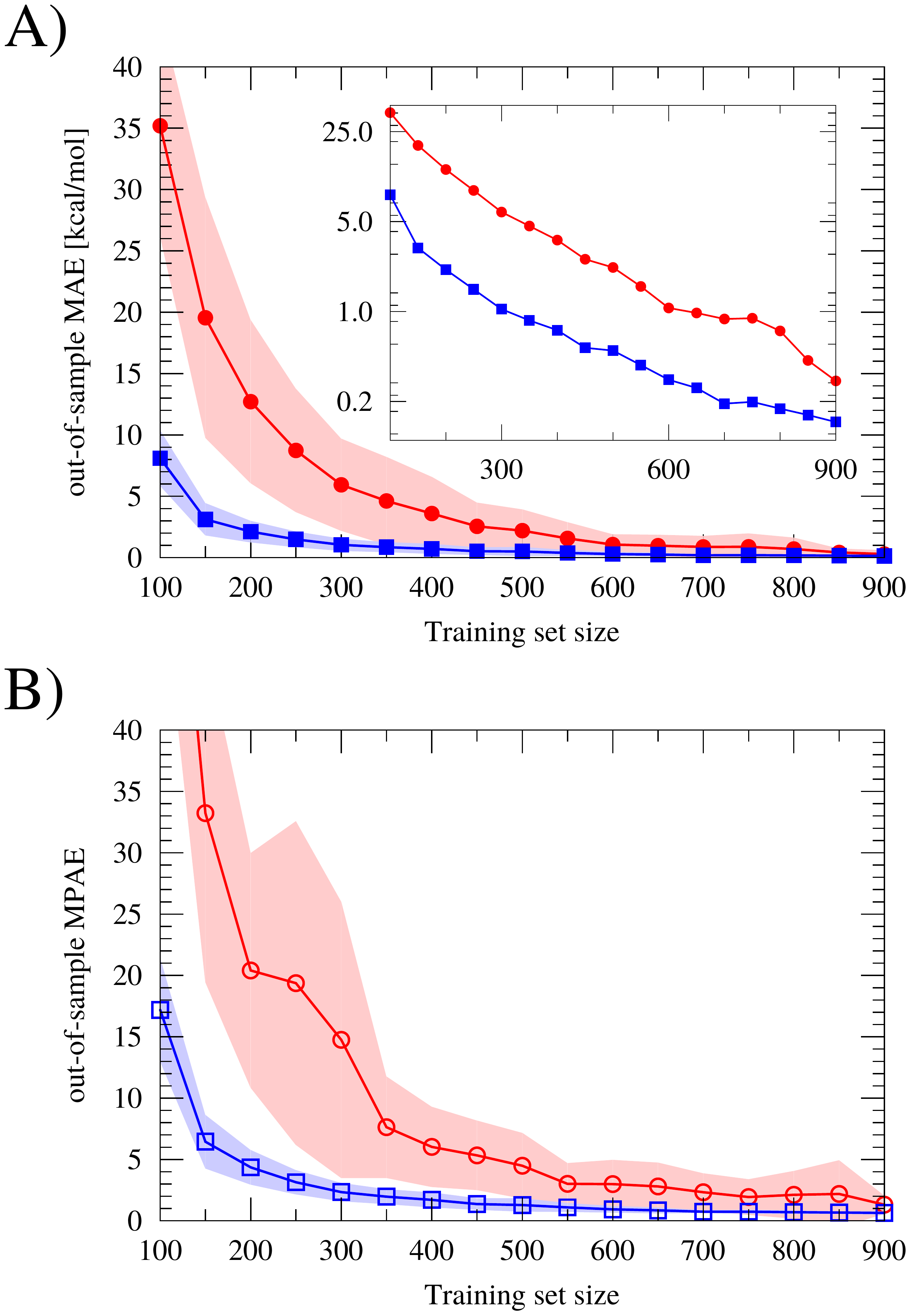}
\caption{                
Out-of-sample errors in correlation energies predicted using Eq.\ref{eq:pred}
as a function of training set size: A) Mean absolute error (MAE) in kcal/mol. The inset shows errors in log scale;
B) Mean percentage absolute error (MPAE).
In both cases, blue squares correspond to prediction errors when the models were
generated after removing the numerical noise in intracules (see text for more details). 
In all cases, the envelope encloses the
standard deviation of the estimate from 100 independent runs.
}
\label{fig:Dia}
\end{figure} 
%===
\begin{figure*}[hpt] 
\centering  
\includegraphics[width=14cm, angle=0.0]{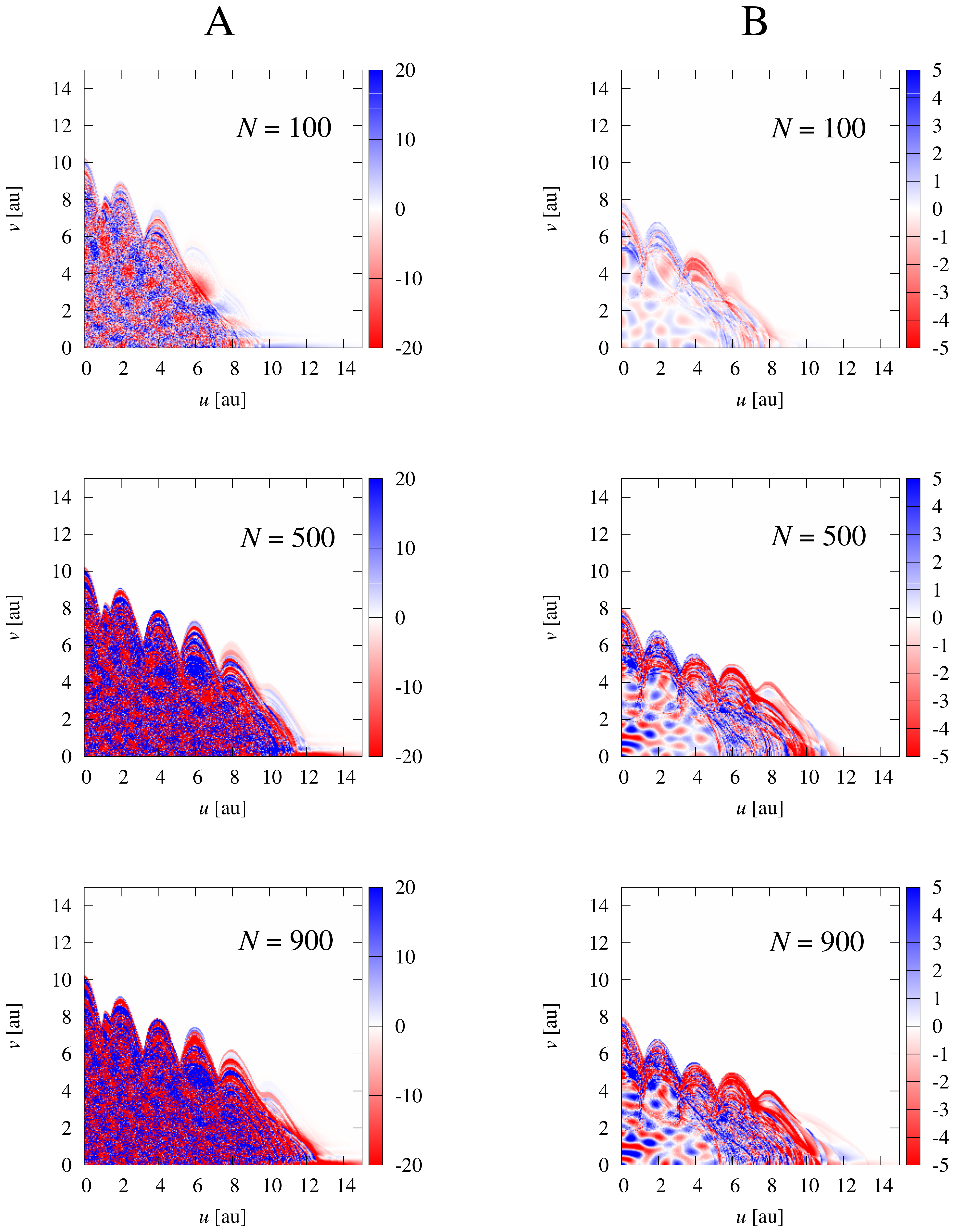}
\caption{ 
ML-predicted intracule kernels, $\mathcal{G}$, for trainingset sizes 100, 500 and 900. 
A) for ML training without noise filtration in $\mathcal{W}$, and B) after filtering noise in
$\mathcal{W}$
}
\label{fig:G}
\end{figure*} 
%===
The system exhibiting the strongest correlation with $E_c=-72.8$ kcal/mol corresponds to a six-well, with five $Z=1$ atomic centers and a terminal atom being $Z=2$. The molecule that exhibits the next strongest correlation is also a six-well, but with all $Z=1$, making it the most delocalized system. To exemplify the trend in $E_c$ across the dataset, for the extreme cases, we have plotted (see Fig.~\ref{fig:psiW}) the two-electron reduced density,  $\rho(x_1,x_2)$, and the corresponding $\mathcal{W}$ computed using Eq.~$\ref{fig:Wuv}$. The figure features the same plots from exact and HF calculations. For the most localized external potential, we find both the correlated and uncorrelated wavefunctions to result in essentially similar $\rho(x_1,x_2)$ and $W(u,v)$, implying a weak post-HF correction that complies with a small $E_c$. On the other hand, for the most delocalized system, the HF density lacks a cusp that is present in the exact density along the $x_1=x_2$ line. These observations are in line with the quantitative trends in $E_c$ noted above.

Based on the trends noted for $\rho(x_1,x_2)$, one can infer the intracule of the single-well, in Fig.~\ref{fig:psiW} to be characteristic of an uncorrelated system. In particular, the corresponding plots imply that an intracule localized at $u=v=0$ can only arise from a wavefunction that intrinsically lacks a Coulomb hole. For the same system, {\it i.e.}, the 1D C$^{4+}$ ion, one could encounter a different $\mathcal{W}$ profile for a different choice of $\alpha$. The intracule of the delocalized system, in contrast, 
exhibits a strong distortion from that of C$^{4+}$. At the HF level, the profile corresponds to an elongated, and somewhat distorted, ellipse lacking a node. The exact intracule, on the other hand, exhibits a non-radial node showing lobes of opposite signs. A lobe centered at $u=0$ and $v=1$ implies that both the electrons tend to move along opposite directions with a relative velocity of 1 atomic unit.

Having discussed the composition of the dataset, and the range of the target property to be modeled, we now discuss the performance of the ML-predicted ${\mathcal{F}}$. Fig.~\ref{fig:Dia} presents the out-of-sample prediction errors of the ML models as a function of the training set size. We present results from two sets of calculations. One, in which ${\widetilde{\mathcal{G}}}$ is obtained by solving Eq.~\ref{eq:dgelsy}. In the other, before training, numerical noise in intracules was filtered. Such noise arises from the numerical integration of Eq.~\ref{fig:Wuv}, resulting in non-zero $\mathcal{W}(u,v)$, of the order of $10^{-6}$, for large $u$ and $v$. As a consequence, the resulting kernels show spurious non-zero values near boundary.  The learning rate in Fig.~\ref{fig:Dia} shows that independent of the noise reduction, with sufficient training, the models forecast reference correlation energies to a mean absolute error (MAE) less than 1 kcal/mol. However, one does note excellent learning rates only after the noise filtration. From the inset of this plot, we find that even for the training set with 300 potentials, the prediction errors drop below the desired threshold. We estimate the uncertainty in the model's performance arising from the training set bias by selecting hundred different random sets. When using noise-free kernels, the prediction errors seemly show sub-kcal/mol standard deviation for training set sizes over 300 while the mean percentage absolute error (MPAE) drops to less than 
1\% for trainingset sizes $>500$. 

Since a analytic form for the ML-intracule-functional is unknown, we could only appreciate the shapes of these functions by plotting them on grids. In Fig.~\ref{fig:G}, we have collected these plots for training set sizes 100, 500 and 900. For all the three choices, we have presented the kernels computed with and without noise filtration. While overall we note the essential shape of the functional to be preserved in all case, we do find the profile to grow denser with more training data. Additionally, we observe de-noising to dampen the $\widetilde{\mathcal{G}}$ while improving its continuity.

\section{Conclusions}
We have introduced a machine learning approach based on the rank-revealing QR decomposition to numerically identify the correlation intracule functional. While using Wigner intracules
derived from Hartree--Fock wavefunctions, the ML-predicted intracule functional yields accurate correlation energies.
Our reference data comprises of 923 1D externals potentials, with 6 atoms (single-wells) and 917 molecules (multi-wells), for which we have computed accurate HF and exact electronic energies using DVR. We have derived an efficient expression to compute the Wigner intracule in 1D, which requires as input the two-electron
wavefunction on a fine grid or as an analytic function. 

Based on the trends in quantum chemistry method development, it would seem that the problem of deriving a closed-form expression for the exact correlation intracule functional
will continue to remain an open challenge, for the foreseeable future. However, 
no severe hurdle seems to be on the way of data-driven modeling of such functionals, at least for 1D 
models of atoms and molecules with two electrons. It remains to be seen if the
approach presented here can be extended to many-electron systems, but still
depending on a reduced Wigner function. Such efforts must also address if Wigner-intracules are $N-$representable, {\emph i.e.,} there is at least one antisymetric, $N$-electron wavefunction of which the Wigner-function is a reduced function\cite{harriman1973}.
With this note, we hope our present study to aid other researchers in the combined application of ML and IFT
to study realistic 3D atoms and molecules.

\section{Acknowledgments}
RB gratefully acknowledges TIFR for Visiting Students’ Research Programme (VSRP) and junior research fellowships. RR thanks TIFR for financial support. The authors thank anonymous referees for thought-provoking comments to a earlier version of the paper. 

\section*{APPENDIX 1: Coulomb interaction in one-dimension}
We know that in 3D, the Coulomb potential ($V(\mathbf{r})$) can be computed as the 
solution of the Poisson equation
\begin{equation}
    \nabla^2 V(\mathbf{r}) = -\frac{\rho(\mathbf{r})}{\epsilon_0}. 
\end{equation}
For a unit positive charge located at $\mathbf{r}'$, the charge-density is a Dirac-delta function, $\rho(\mathbf{r})=\delta(\mathbf{r}-\mathbf{r}')$. 
The Poisson equation when solved with the appropriate boundary condition,
$\underset{|{\mathbf{r}| \rightarrow \infty}}{\rm lim} V(\mathbf{r}) = 0$,
results in 
\begin{equation}
    \nabla^2 V(\mathbf{r}) = -\frac{\delta(\mathbf{r} - \mathbf{r}') }{\epsilon_0};\quad
    V(\mathbf{r}) = -\frac{1}{4 \pi \epsilon_0} \frac{1}{|\mathbf{r} - \mathbf{r}'|}.
\end{equation}
The solution may be checked using the Green's function
\begin{eqnarray}
G(\zeta)=\frac{1}{4\pi |\zeta|};  \quad  \nabla^2 G(\zeta) =   \delta(\zeta).
\end{eqnarray}

In comparison, in 1D, for a unit positive charge located at $x'$, Poisson equation takes the form
\begin{equation}
    d^2_x V(x) = \frac{-\delta(x-x')}{\epsilon_0},
\end{equation}
the boundary condition being 
\begin{eqnarray}
d_x V(x)|_{x^{\prime}+\Delta x} = -d_x V(x)|_{x^{\prime}-\Delta x} 
\end{eqnarray}
for any positive $\Delta x$. The solution is then the 1D Coulomb potential which
is linear in $x-x'$
\begin{equation}
 V(x)= -\frac{1}{2 \epsilon_0}|x-x'|.
\end{equation}
This potential gives rise to a uniform electric field up to a change in sign, $E(x>x')=-1/(2\epsilon_0);\, E(x<x')=1/(2\epsilon_0)$.

\section*{APPENDIX 2: A note on hard-Coulomb interactions in one-dimension}
\paragraph*{\bf Classical case:}Let us consider a single-well (atomic) system with two electrons. For convenience, let the nucleus be fixed at the origin, $x=0$. The hard-Coulomb external and electron-repulsion potentials are $V_{\rm ext}(x_1, x_2)=-Z(1/|x_1|+1/|x_2|)$ and $V_{\rm ee}(x_1,x_2)=1/|x_1-x_2|$, where $x_1$ and $x_2$ are the coordinates of the two electrons and $Z$ is the atomic number. In classical mechanics, such a system will reach equilibrium when the net-force on every electron $F_{{\rm ext}, i} + F_{{\rm ee}, i}= 0$.
Using symmetry arguments, it can be shown that this  condition is reached only when the particles are equally displaced from the nucleus, $x_1=-x_2$. For electron-1 ($i=1$) the force-balance 
criterion leads to the relation
\begin{eqnarray}
\frac{Z}{|x_1|^2} - \frac{1}{|x_1-x_2|^2} & = & 0 \nonumber \\
% \Rightarrow  \frac{Z}{|x_1|^2} - \frac{1}{4|x_1|^2} & = & 0  \nonumber \\
\Rightarrow  \left( Z-1/4 \right) \frac{1}{|x_1|^2} & = & 0  
 \end{eqnarray}
 For any integer nuclear charge, $Z$, the system cannot be in equilibrium for finite $x_1$. So, starting with any finite electronic positions, the system will approach the least-energy state corresponding to $x_1=x_2=0$,
 with $V_{\rm ext}+V_{\rm ee}\rightarrow-\infty$. 

\paragraph*{\bf Quantum mechanical case:} In the quantum mechanical version of the same system, let us start with a (normalized) singlet trial-wavefunction, $\psi(x_1,x_2)$, satisfying the following two conditions
\begin{enumerate}
    \item Kinetic energy and electron-electron repulsion expectation values are finite; which also accounts for the Coulomb-hole condition $|\psi(x_1,x_1)|^2=0\,\forall\,x_1\in \mathbb{R}$. 
    \item For one electron at the origin, the wavefunction does not vanish ($\psi(0, x_2)\ne0$) in the neighborhood of at least one non-trivial value of $x_2=y\ne0$. In other words, the minimal value of $|\psi(x_1,x_2)|^2=\beta$ is non-vanishing 
    in the domain $(-\epsilon  ,  y - \epsilon) < (x_1, x_2) < (+\epsilon , y + \epsilon) $ for a small positive $\epsilon$. 
\end{enumerate}
An example function satisfying both these conditions is 
\begin{eqnarray}
g(x_1,x_2) = N  e^{a_1 (x_1 - x_2)^2 + a_2  (x_1 + x_2)^2} (x_1 - x_2)^2, 
\end{eqnarray}
where $a_1,a_2\in  \mathbb{R}$ and $N$ is the appropriate normalization factor.

While the second of the aforestated conditions ensures that the minimal value of $|\psi(x_1,x_2)|^2=\beta$ is non-vanishing around the point $(0,y)$, it is easy to see that in the same domain, the expectation value of $\hat{V}_{n}$ diverges:
\begin{eqnarray} 
 \langle \hat{V}_{ne} \rangle  &= & -Z\iint_{-\infty}^{\infty} dx_1 dx_2 |\psi(x_1,x_2)|^2 \Big( \frac{1}{|x_1|} + \frac{1}{|x_2|}  \Big) \nonumber \\
 &=&  -2 Z \iint_{-\infty}^{\infty} dx_1 dx_2 |\psi(x_1,x_2)|^2 \frac{1}{|x_1|} \nonumber \\
 % & \geq &    2 \int_{y -\epsilon}^{y+ \epsilon} dx_2 \int_{-\epsilon}^{\epsilon} dx_1 |\psi(x_1,x_2)|^2 \Big( \frac{1}{|x_1|}  \Big) \nonumber \\
 & \leq &  -2 Z \int_{y -\epsilon}^{y+ \epsilon} dx_2 \int_{-\epsilon}^{\epsilon} dx_1 \beta \frac{1}{|x_1|} \nonumber \\
 &=& -4 Z \epsilon \beta  \int_{-\epsilon}^{+\epsilon} dx_1  \frac{1}{|x_1|} = -8 Z \epsilon \beta  [ \ln(|\epsilon| - \ln(0) ] \nonumber \\
 %&=& -8 Z \epsilon \beta  ( \ln(|\epsilon| - \ln(0) ) \nonumber \\
 & \rightarrow & -\infty 
\end{eqnarray}
Thus, for the trial-wavefunction chosen, $\langle \hat{H} \rangle \rightarrow - \infty$. It is now 
fairly straightforward to apply variational principle and show that any other choice of trial-wavefunction $\psi$
will satisfy $\langle \psi | \hat{H} | \psi \rangle \geq E_{g}$, where $E_g$ is the ground state energy.
Hence, the upper bound for the ground state energy of a hard-Coulomb two-electron system always diverges towards negative infinity.

\section*{References}

\bibliography{lit} 

\end{document}